\documentclass[aps,graphicx,twocolumn]{revtex4}
\usepackage{amssymb}
\usepackage{txfonts}
\usepackage{graphicx}

\begin{document}

\title{Single-photon-assisted entanglement concentration of a multi-photon system in a partially entangled W
state with  weak cross-Kerr nonlinearity\footnote{Published in J. Opt. Soc. Am. B  \textbf{29}, 1399 (2012)}}

\author{ Fang-Fang Du, Tao Li,  Bao-Cang Ren,  Hai-Rui Wei, and Fu-Guo Deng\footnote{Corresponding author. Email address:
fgdeng@bnu.edu.cn} }
\address{ Department of Physics, Applied Optics Beijing Area Major Laboratory, Beijing Normal University, Beijing 100875, China}
\date{\today }

\begin{abstract}
We propose a  nonlocal entanglement concentration protocol (ECP) for
$N$-photon systems in a partially entangled W state, resorting to
some ancillary single photons and the parity-check measurement based
on cross-Kerr nonlinearity. One party in quantum communication first
performs a parity-check measurement on her photon in an $N$-photon
system and an ancillary photon, and then she picks up the
even-parity instance for obtaining the standard W state. When she
obtains an odd-parity instance, the system is in a less-entanglement
state and it is the resource in the next round of entanglement
concentration. By iterating the entanglement concentration process
several times, the present ECP has the total success probability
approaching to the limit in theory. The present ECP has the
advantage of a high success probability.  Moreover, the present ECP
requires only the $N$-photon system itself and some ancillary single
photons, not two copies of the systems, which decreases the
difficulty of its implementation largely in experiment.  It maybe
have good applications in quantum communication in future.\\

\end{abstract}


 \maketitle

\section{introduction}

Entanglement is a key important resource in quantum information and
quantum computation \cite{book}. The advantage of quantum computer,
the powerful computation, comes from  multipartite entanglement,
compared with classical computer.  Also, entanglement is used as the
information carries in quantum communication, such as quantum key
distribution (QKD) \cite{Ekert91,BBM92,Longliu,denglong,lidengzhou},
quantum teleportation \cite{teleportation}, quantum dense coding
\cite{densecoding,super2}, quantum secret sharing
\cite{QSS1,QSS2,QSS3,QSS4,QSS5,QSS6,QSS7,QSS8}, quantum state
sharing \cite{QSTS1,QSTS2,QSTS3,QSTS4,QSTS5}, controlled
teleportation
\cite{cteleportation1,cteleportation2,cteleportation3}, and so on.
In a long-distance quantum communication, entanglement is used to
construct quantum repeaters. However, entanglement  is fragile to
channel noise. In a practical transmission or the process for
storing an entangled quantum system, it inevitably suffers from
channel noise and its environment. The noise will make the system
decoherent, which will decrease the security of QKD protocols and
the fidelity of quantum teleportation and dense coding. There are
some interesting ways for dealing with the issue of decoherence in
quantum communication, such as decoherence-free subspaces \cite{r34,
r35, r36, r37}, faithful qubit distribution \cite{r40, r41},
faithful qubit transmission \cite{r42}, error-rejecting codes
\cite{r43}, faithful entanglement distribution \cite{r44}, and so
on. Most of them are very useful for overcoming a collective noise
by encoding a logical qubit with several physical qubits. There is a
fundamental hypothesis that the noise is a collective one. These
methods are used to deal with the photon systems before they are
transmitted over a noisy channel.

Entanglement purification and entanglement concentration are two
interesting quantum techniques with which the users can obtain some
high-fidelity entangled photon systems after they are transmitted
over a noisy channel or stored in a practical environment and they
are in a less-entanglement state. In detail, entanglement
purification is used to extract some high-fidelity entangled systems
from a less-entangled ensemble in a mixed state
\cite{Bennett1,Deutsch,Pan1,Simon,shengpra,shengpratwostep,lixhdepp,shengpraonestep,dengmdepp,wangcpra,wangcqic,dengEMEPP}.
Entanglement concentration  is used to obtain a subset of photon
systems in a maximally entangled state from a set of systems in a
partially entangled pure state. Although entanglement purification
is more general than entanglement concentration in the practical
applications because an entangled photon system is usually in a
mixed entangled state after it is transmitted over a noisy channel,
entanglement concentration is entanglement concentration is more
efficient for the two remote parties in quantum communication, say
the sender Alice and the receiver Bob, to distill some maximally
entangled systems from an ensemble in a less-entangled pure state
because entanglement purification  should consume a great deal of
quantum resource for  improving the fidelity of systems in a mixed
entangled state, not obtain a maximally entangled state directly.
Entanglement concentration is useful in some particular cases, such
as the decoherence of entanglement arising from the storage process
or the imperfect entanglement source.

Since Bennett \emph{et al.} \cite{Bennett2} proposed the original
entanglement concentration protocol (ECP) in 1996, there have been
some interesting and typical ECPs for photon systems
\cite{Bennett2,Yamamoto,zhao1,shengpraecp,shengsingle,swapping1,swapping2,shengecp,dengecppra}
and atom systems \cite{caozl1,caozl2}. For example, Bose \emph{et
al.} \cite{swapping1} proposed an ECP based on entanglement swapping
in 1999. Subsequently,  Shi \emph{et al.} \cite{swapping2} presented
an ECP based on a collection unitary evolution on a qubit in a
multi-qubit system and an ancillary qubit. In 2001, an ECP based on
polarizing beam splitters (PBSs) was proposed \cite{Yamamoto,zhao1}.
In 2008, Sheng, Deng and Zhou proposed an interesting ECP
\cite{shengpraecp}with cross-Kerr nonlinearities. In 2010, they
presented the first single-photon ECP \cite{shengsingle} with
cross-Kerr nonlinearities.  In 2012, Sheng \emph{et al.}
\cite{shengecp} proposed a single-photon-assisted ECP for partially
entangled multi-photon systems. Recently,   an optimal nonlocal
multipartite ECP for photon systems in a partially entangled
Bell-type state is proposed \cite{dengecppra}, resorting to a
parity-check measurement on one photon in the system and an
ancillary single photon and the projection measurement on the
ancillary  photon with cross-Kerr nonlinearities.

Although there exist some interesting ECPs, most of them are used to
distill some maximally entangled Bell  states or
Greenberger-Horne-Zeilinger (GHZ) states. There are few schemes for
concentrating the non-maximally entangled pure W-class states. In
essence, W state are inequivalent to the GHZ states as they cannot
be converted into each other  under stochastic local operations and
classical communication (SLOCC). Moreover,  a W state is more robust
than a GHZ state against the loss of one or two photons.  Therefore,
it is interesting to discuss the entanglement concentration on the
partially entangled W state. By far, there are three ECPs for photon
systems in a partially entangled W state \cite{W0,W1,W2}.  The first
one is a linear optical scheme for entanglement concentration of two
known  partially entangled three-photon W states \cite{W0}. The
second one is linear-optics-based entanglement concentration of
unknown partially entangled three-photon W states \cite{W1}. It is
proposed by  Wang, Zhang, and Yeon \cite{W1} in 2010. In 2011, Xiong
and  Ye \cite{W2} proposed another ECP for a partially entangled W
state with cross-Kerr nonlinearity. Both these two interesting ECPs
are used to deal with an unknown multi-photon W-class state. There
is no ECP for a  known multi-photon W-class state.

In this paper, we proposed an  nonlocal ECP for $N$-photon systems
in a known partially entangled pure W state, resorting to ancillary
single photons and the parity-check measurement based on cross-Kerr
nonlinearity. It does not depend on two copies of $N$-photon systems
in a partially entangled W-class state in each round of
concentration, just each system itself and some ancillary single
photons, which makes it far different from other ECP for W states
\cite{W1,W2}. In the present ECP, only one of the parties in quantum
communication, say Alice, first operates her photon and the
ancillary single photons for concentrating the entanglement of an
$N$-photon system and then tells the others to retain or discard the
 system,  which greatly simplifies
the complication of classical communication as others require all
the parties operate the entanglement process in the same way,
similar to the works for a Bell-type state
\cite{shengecp,dengecppra}. Moreover, the present ECP has a higher
total success probability approaching to the limit in theory by
iterating the entanglement concentration process several times. All
these advantages make our ECP more feasible than others. It maybe
have good applications in quantum communication in future.

\section{Entanglement concentration of partially entangled three-photon W states}

\begin{figure}[!h]
\begin{center}
\includegraphics[width=8cm,angle=0]{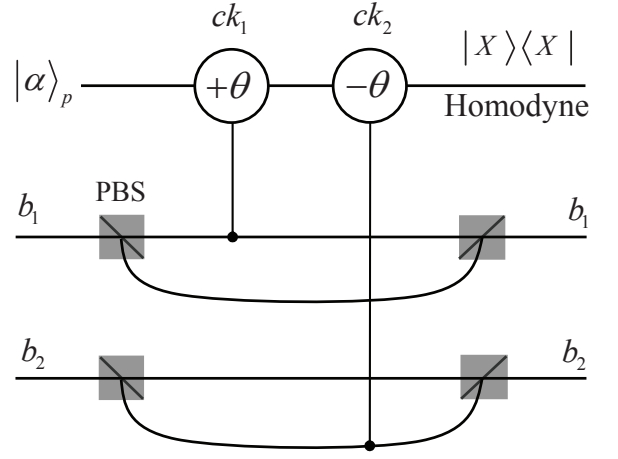}
\caption{The principle of a parity-check detector (PCD) on the
polarizations of two photons, the same as that in
Ref.\cite{dengEMEPP,dengecppra}. PBS represents a polarizing beam
splitter which transmits the photon in the horizontal polarization
$\vert H\rangle$ and reflects the photon in the vertical
polarization $\vert V \rangle$. $ck_1$ and $ck_2$ represent two
cross-Kerr nonlinearities which will lead to the phase shift
$+\theta$ and  $-\theta$  when there is a photon passing through the
media, respectively. $|X\rangle\langle X|$ represents an X
quadrature measurement  with which one can not distinguish the the
states $\vert \alpha e^{\pm i\theta} \rangle_p$ \cite{QND1,QND3}.}
\label{fig1_QND}
\end{center}
\end{figure}

Before we discuss the principle of our ECP for a partially entangled
three-photon W states, we first introduce the principle of a
parity-check detector (PCD) on the polarization states of two
photons with cross-Kerr nonlinearity.  In fact, the principle of the
PCD here is similar to those in
Refs.\cite{QND1,dengEMEPP,dengecppra}. In detail, the Hamiltonian of
a cross-Kerr nonlinearity can be written as follows \cite{QND1}:
\begin{eqnarray}
H_{ck}=\hbar\chi a^{+}_{s}a_{s}a^{+}_{p}a_{p}
\end{eqnarray}
where $a^{+}_{s}$ and $a^{+}_{p}$ are the creation operations, and
$a_{s}$ and $a_{p}$ are the destruction operations. The subscripts
$s$ and $p$ represent the signal light and the probe light,
respectively.  $\chi$ is the coupling strength of the cross-Kerr
nonlinearity. If a signal light in the state
$|\Psi\rangle_s=c_{0}|0\rangle_{s}+c_{1}|1\rangle_{s}$
($|0\rangle_{s}$ and $|1\rangle_{s}$ denote that there are no photon
and one photon respectively in this state) and a coherent probe beam
in the state $|\alpha\rangle_p$ couple with a cross-Kerr
nonlinearity medium, the evolution of the whole system  can be
described as \cite{QND1,dengEMEPP,dengecppra}:
\begin{eqnarray}
U_{ck}|\Psi\rangle_{s}|\alpha\rangle_{p}&=&
e^{iH_{ck}t/\hbar}[c_{0}|0\rangle_{s}+c_{1}
|1\rangle_{s}]|\alpha\rangle_{p} \nonumber\\
&=& c_{0}|0\rangle_{s}|\alpha\rangle_{p}+c_{1}|1\rangle_{s}| \alpha
e^{i\theta}\rangle_{p},
\end{eqnarray}
where $\theta=\chi t$ is the phase shift of the probe beam, which
depends on  the interaction time  $t$ and the  coupling strength
$\chi$. That is, the coherent beam $P$ picks up a phase shift
$\theta$ directly proportional to the number of the photons in the
signal light in the Fock state $|\Psi\rangle_s$. Based on this
feature of a cross-Kerr nonlinearity, the principle of our PCD is
shown in Fig.\ref{fig1_QND}, similar to those in Refs.
\cite{QND1,dengEMEPP,dengecppra}. Here $\vert X\rangle\langle
X\vert$ represents an X quadrature measurement with which one can
not distinguish the the states $\vert \alpha e^{\pm i\theta}
\rangle_p$  \cite{QND1,QND3}. With the two cross-Kerr nonlinearities
$ck_1$ and $ck_2$,  one can distinguish the superpositions and
mixtures of the polarization states $|HH\rangle$ and $|VV\rangle$
from $|HV\rangle$ and $|VH\rangle$ based on the different phase
shifts. That is,  the probe beam $\vert \alpha \rangle_p$ will pick
up a phase shift $\theta$ if the two photons is in the state
$|HH\rangle_{b_1b_2}$ or $|VV\rangle_{b_1b_2}$, while it picks up a
phase shift $0$ when the two photons is in the state
$|VH\rangle_{b_1b_2}$ or $|HV\rangle_{b_1b_2}$. In other words, when
the parity of the two photons is even, the coherent beam will pick
up a phase shift $\theta$; otherwise it will pick up  a phase shift
$0$. By detecting the phase shift of the probe beam, one can
determine that the parity of the two photons is even or odd.

\begin{figure}[!h]
\begin{center}
\includegraphics[width=8cm,angle=0]{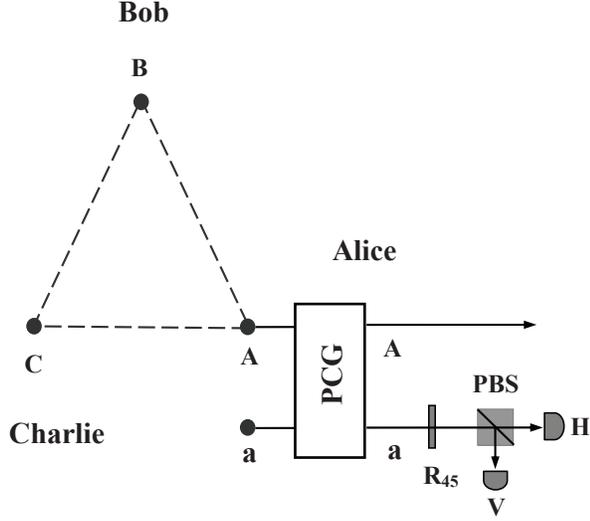}
\caption{The principle of the present ECP for three-photon W states
with ancillary single photons and the parity-check detector (PCD)
based on cross-Kerr nonlinearity. $R_{45}$ represents a Hadamard
operation on the polarization of the ancillary single photon $a$. H
and V represent the horizontal polarization state of photons $\vert
H\rangle$ and  the vertical polarization state $\vert V \rangle$,
respectively.} \label{fig1_w}
\end{center}
\end{figure}

Let us assume that the three-photon system composed of the three
photons $ABC$ is in the following partially   entangled W-class
polarization states:
\begin{eqnarray}
|\varphi\rangle_{CBA} &=&
\alpha|H\rangle_{C}|H\rangle_{B}|V\rangle_{A} +
\beta(|H\rangle_{C}|V\rangle_{B}|H\rangle_{A}\nonumber\\
 &+&
|V\rangle_{C}|H\rangle_{B}|H\rangle_{A}),\label{state01}
\end{eqnarray}
where the subscripts $A$, $B$,  and $C$ represent the three photons
belonging to the three remote parties in quantum communication, say
Alice, Bob, and Charlie. Different from those ECPs for W states
\cite{W1,W2}, here $\alpha$ and $\beta$ are two known real numbers
and satisfy the relation
\begin{eqnarray}
\alpha^{2}+ 2\beta^{2}=1.
\end{eqnarray}
Certainly, in a practical application, it is not difficult for the
parties to obtain information about the parameters  $\alpha$ and
$\beta$ by detecting a subset of three-photon systems, similar to
the case for Bell-type states \cite{dengecppra}.

The principle of our ECP is shown in Fig.\ref{fig1_w}. In the
process of concentrating a three-photon system, Alice prepares an
ancillary photon $a$. It is in the polarization state
$\vert\varphi\rangle_a$. Here
\begin{eqnarray}
|\varphi\rangle_{a}=\frac{1}{\sqrt{\alpha^2 +
\beta^2}}(\alpha|H\rangle + \beta|V\rangle)_{a}.\label{state02}
\end{eqnarray}
Before Alice preforms a parity-check measurement on her photon $A$
and the ancillary photon $a$, the composite system composed of the
four photons $CBAa$ is in the state
\begin{eqnarray}
|\Phi\rangle_{CBAa} &=& |\varphi\rangle_{CBA}\otimes
|\varphi\rangle_{a}\nonumber\\
&=&\frac{1}{\sqrt{\alpha^2 + \beta^2}}\{ \alpha\beta
[|H\rangle_C\vert H\rangle_B \vert V\rangle_A \vert
V\rangle_a + (\vert H\rangle_C \vert V\rangle_B \nonumber\\
&+&  \vert V\rangle_C \vert H\rangle_B)\vert H\rangle_A\vert
H\rangle_a] + \alpha^2 |H\rangle_{C}|H\rangle_{B}|V\rangle_{A}\vert
H\rangle_a   \nonumber\\
&+& \beta^2 (|H\rangle_{C}|V\rangle_{B} +
|V\rangle_{C}|H\rangle_{B})|H\rangle_{A}|V\rangle_{a}  \}.
\label{state03}
\end{eqnarray}
With the parity-check measurement on the photons $A$ and $a$, Alice
can divide the state of the four-photon system $CBAa$ into two
classes. In the first one, it is in a  state in which each item has
the same parameter, that is,
\begin{eqnarray}
|\Psi_1\rangle_{CBAa} &=& \frac{1}{\sqrt{3}}(|H\rangle_C\vert
H\rangle_B \vert V\rangle_A \vert V\rangle_a + (\vert H\rangle_C
\vert V\rangle_B  \nonumber\\
&+& \vert V\rangle_C \vert H\rangle_B)\vert H\rangle_A\vert
H\rangle_a). \label{state04}
\end{eqnarray}
In the second one, the system is in a state with less entanglement
and different parameters, that is,
\begin{eqnarray}
|\Psi'_1\rangle_{CBAa} &=& \frac{1}{\sqrt{\alpha^2 +
2\beta^2}}[\alpha^2 |H\rangle_{C}|H\rangle_{B}|V\rangle_{A}\vert
H\rangle_a \nonumber\\
&+&   \beta^2 (|H\rangle_{C}|V\rangle_{B} +
|V\rangle_{C}|H\rangle_{B})|H\rangle_{A}|V\rangle_{a} ].
\label{state05}
\end{eqnarray}
In the fact, in the first class, Alice obtains an even parity when
she performs a parity-check measurement on the photon $A$ and the
ancillary $a$. The state $|\Psi_1\rangle_{CBAa}$ corresponds to the
parameter $\alpha\beta$ in Eq.(\ref{state03}). In the second class,
Alice obtains an odd parity, which leads to the state
$|\Psi_1'\rangle_{CBAa}$. The probability that Alice obtains an even
parity when she measures the two photons $A$ and $a$ is
\begin{eqnarray}
P_1= \frac{3\alpha^2\beta^2}{  \alpha^2 + \beta^2 }. \label{P}
\end{eqnarray}
The probability that Alice obtains an odd parity   is
\begin{eqnarray}
P'_1= \frac{\alpha^4 + 2\beta^4}{  \alpha^2 + \beta^2 }. \label{P}
\end{eqnarray}

Alice can measure the ancillary photon $a$ for obtaining the
standard three-photon W state from the four-photon state
$|\Psi_1\rangle_{CBAa}$ with the basis $X$ (i.e.,  $ \{ \vert \pm
x\rangle =\frac{1}{\sqrt{2}}(\vert H\rangle \pm \vert V\rangle)
\}$). When she obtain the state $\vert +x\rangle_a$, the
three-photon system is in the standard W state $\vert W^+_3\rangle$.
Here
\begin{eqnarray}
\vert W^+_3\rangle_{CBA} = \frac{1}{\sqrt{3}}(|H\rangle_C\vert
H\rangle_B \vert V\rangle_A + (\vert H\rangle_C \vert V\rangle_B +
\vert V\rangle_C \vert H\rangle_B)\vert H\rangle_A).
\nonumber\\
\label{state06}
\end{eqnarray}
When she obtain the state $\vert -x\rangle_a$, the three-photon
system is in another standard W stat
\begin{eqnarray}
\vert W^-_3\rangle_{CBA} = \frac{1}{\sqrt{3}}(|H\rangle_C\vert
H\rangle_B \vert V\rangle_A - (\vert H\rangle_C \vert V\rangle_B +
\vert V\rangle_C \vert H\rangle_B)\vert H\rangle_A).
\nonumber\\
\label{state06}
\end{eqnarray}
Alice can transform the state $\vert W^-_3\rangle_a$ into the state
$\vert W^+_3\rangle$ by performing a phase-flip operation
$\sigma_z=\vert H\rangle\langle H\vert - \vert V\rangle\langle
V\vert$ on her photon $A$.

As for the less-entanglement state $|\Psi'_1\rangle_{CBAa}$, Alice
can  measure the ancillary photon $a$ with the basis $X$ to
transform it into a three-photon state with less entanglement. That
is,
\begin{eqnarray}
|\Psi'_2\rangle_{CBA} &=& \frac{\alpha^2}{\sqrt{\alpha^2 +
2\beta^2}} |H\rangle_{C}|H\rangle_{B}|V\rangle_{A}  \nonumber\\
&+&    \frac{\beta^2}{\sqrt{\alpha^2 + 2\beta^2}}
(|H\rangle_{C}|V\rangle_{B} +
|V\rangle_{C}|H\rangle_{B})|H\rangle_{A}  . \label{state07}
\end{eqnarray}
In detail, when Alice obtains the state $\vert +x\rangle_a$, the
three-photon system is in the state $|\Psi'_2\rangle_{CBA}$. When
Alice obtains the state  $\vert -x\rangle_a$, she need only perform
a phase-flip operation on her photon $A$ and she will obtain the
state  $|\Psi'_2\rangle_{CBA}$.

It is not difficult to find that the state  $|\Psi'_2\rangle_{CBA}$
shown in Eq.(\ref{state07}) has the same form as the state $\vert
\varphi\rangle_{CBA}$ shown in Eq.(\ref{state01}) but different
parameters. We need only replace the parameters $\alpha$ and $\beta$
in  Eq.(\ref{state01}) with the parameters $\alpha'\equiv
\frac{\alpha^2}{\sqrt{\alpha^2 + 2\beta^2}} $ and $ \beta'\equiv
\frac{\beta^2}{\sqrt{\alpha^2 + 2\beta^2}}$, respectively. That is,
Alice can also concentrate the state  $|\Psi'_2\rangle_{CBA}$ as the
same as the state $\vert \varphi\rangle_{CBA}$. The probability that
Alice, Bob, and Charlie obtain the standard three-photon W state
from each system in the stat  $|\Psi'_2\rangle_{CBA}$ is
\begin{eqnarray}
P_2&=&  \frac{  3\cdot \frac{\alpha^4}{\alpha^4+2\beta^4}\cdot
\frac{\beta^4}{\alpha^4+2\beta^4}}{\frac{\alpha^4}{\alpha^4+2\beta^4}
+ \frac{\beta^4}{\alpha^4+2\beta^4}} =
\frac{3\alpha^4\beta^4}{(\alpha^4 + \beta^4)(\alpha^4 + 2\beta^4)}.
\end{eqnarray}
Certainly, the probability that Alice, Bob, and Charlie obtain the
three-photon state with less entanglement from each system in the
stat $|\Psi'_2\rangle_{CBA}$ becomes
\begin{eqnarray}
P'_2&=&  \frac{  (\frac{\alpha^4}{\alpha^4+2\beta^4})^2 + 2\cdot
(\frac{\beta^4}{\alpha^4+2\beta^4})^2}{\frac{\alpha^4}{\alpha^4+2\beta^4}
+ \frac{\beta^4}{\alpha^4+2\beta^4}} = \frac{\alpha^8 +
2\beta^8}{(\alpha^4 + \beta^4)(\alpha^4 + 2\beta^4)}.\nonumber\\
\end{eqnarray}
After Alice performs the entanglement concentration process for $n$
times, the total probability that Alice, Bob, and Charlie obtain the
standard three-photon W state $\vert W^+_3\rangle_{CBA}$ is
\begin{eqnarray}
P(n) &=& P_1 + P'_1P_2+ + P'_1P'_2P_3 + \cdots + P'_1P'_2\cdots
P'_{n-1}P_n\nonumber\\
&=& 3[ \frac{\alpha^2\beta^2}{\alpha^2 + \beta^2}+
\frac{\alpha^4\beta^4}{(\alpha^2 + \beta^2)(\alpha^4 + \beta^4)} \nonumber\\
&+& \frac{\alpha^{8}\beta^{8}}{(\alpha^2 + \beta^2)(\alpha^4 +
\beta^4)(\alpha^8 + \beta^8)} +
\cdots\nonumber\\
&+& \frac{\alpha^{2^n}\beta^{2^n}}{(\alpha^2 + \beta^2)(\alpha^4 +
\beta^4)\cdots(\alpha^{2^n}+\beta^{2^n})}]. \label{pn}
\end{eqnarray}

\begin{figure}[!h]
\begin{center}
\includegraphics[width=8cm,angle=0]{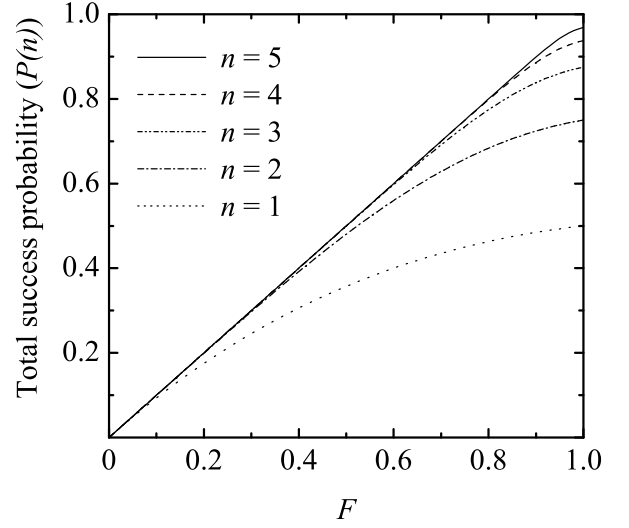}
\caption{ The relation between the total success probability $P(n)$
and the parameter $F=3|\alpha|^2$ when $|\alpha|^2 \leq |\beta|^2$
for the cases  $n=$ 1 (dot curve), 2 (dash-dot curve), 3
(dash-dot-dot curve), 4 (dash curve),  and 5 (solid curve),
respectively. }\label{fig3}
\end{center}
\end{figure}

Let us assume that the parameter $|\alpha|^2 \leq |\beta|^2$. One
can see that  the maximal success probability that Alice, Bob, and
Charlie can distill a standard W state from the partially entangled
state $|\varphi\rangle_{CBA} =
\alpha|H\rangle_{C}|H\rangle_{B}|V\rangle_{A} +
\beta(|H\rangle_{C}|V\rangle_{B}|H\rangle_{A}+
|V\rangle_{C}|H\rangle_{B}|H\rangle_{A})$ is $3|\alpha|^2$ and
$|\alpha|^2 \in [0, 1/3]$. Let us assume $F=3|\alpha|^2$. The
relation between the total probability  $P(n)$ and $F$ is shown in
Fig.\ref{fig3}. Generally, when Alice repeats her entanglement
concentration 5 times, the total success probability $P(n)$
approaches to the parameter $F$, the limit in theory. For a
partially entangled W-class state with less entanglement, Alice need
only iterate the process for 2 or 3 times for obtaining the total
success probability approaching to the limit.

\section{Entanglement concentration of  partially entangled $N$-photon W states}

In principle, it is not difficult to generalize our ECP for
partially entangled $N$-photon W states. Let us assume that there is
a partially entangled $N$-photon W-class state
\begin{eqnarray}
|\varphi\rangle_{ABC\cdots Z} &=& \alpha_1\vert H\rangle_Z\cdots
\vert
H\rangle_C\vert H\rangle_B |V\rangle_A \nonumber\\
&+& \beta_1(\vert H \rangle_Z \cdots\vert H\rangle_C\vert
V\rangle_B |H\rangle_A \nonumber\\
&+&\vert H
\rangle_Z \cdots \vert V\rangle_C\vert H\rangle_B   |H\rangle_A+ \cdots \nonumber\\
&+& \vert V \rangle_Z \cdots \vert H\rangle_C \vert H\rangle_B |H\rangle_A).\label{state08}
\end{eqnarray}
The subscript $A$, $B$, $C$, $\dots$, and $Z$ represent the photons
in W-class states shared by   Alice, Bob, Charlie, $\dots$, and
Zach, respectively. Here, the parameters $\alpha_1$ and $\beta_1$
satisfy the following  relation
\begin{eqnarray}
\alpha_1^{2}+ (N-1)\beta_1^{2}=1.
\end{eqnarray}

For obtain a standard $N$-photon W state from each system in the
state $|\varphi\rangle_{ABC\cdots Z}$, Alice prepares an ancillary
photon $a_1$ in the stat $\vert
\varphi\rangle_{a_1}=\frac{1}{\sqrt{\alpha_1^2 +
\beta_1^2}}(\alpha_1 \vert H\rangle + \beta_1 \vert \vert
V\rangle)_{a_1}$, similar to the case in the entanglement
concentration of a three-photon system. Then the state of the
composite system can be written as
\begin{eqnarray}
|\Phi\rangle_{Z\cdots CBAa_1} &=& |\varphi\rangle_{Z\cdots
CBA}\otimes
|\varphi\rangle_{a_1}\nonumber\\
&=&\frac{1}{\sqrt{\alpha_1^2 + \beta_1^2}}\{ \alpha_1\beta_1 [\vert
H\rangle_Z\cdots|H\rangle_C\vert H\rangle_B \vert V\rangle_A \vert
V\rangle_{a_1} \nonumber\\
&+&  (\vert H\rangle_Z\cdots \vert H\rangle_C \vert V\rangle_B +
\vert H\rangle_Z\cdots \vert V\rangle_C \vert H\rangle_B \nonumber\\
&+& \cdots + \vert V\rangle_Z\cdots \vert H\rangle_C \vert
H\rangle_B)\vert H\rangle_A\vert H\rangle_{a_1}]
\nonumber\\
&+& \alpha_1^2 \vert H\rangle_Z\cdots
|H\rangle_{C}|H\rangle_{B}|V\rangle_{A}\vert
H\rangle_{a_1}   \nonumber\\
&+& \beta_1^2 (\vert H\rangle_Z\cdots|H\rangle_{C}|V\rangle_{B} +
\vert
H\rangle_Z\cdots|V\rangle_{C}|H\rangle_{B} \nonumber\\
&+& \cdots + \vert
V\rangle_Z\cdots|H\rangle_{C}|H\rangle_{B})|H\rangle_{A}|V\rangle_{a_1}
\}. \label{state09}
\end{eqnarray}

If the parity of the two photons $A$ and $a$ is even, the
$(N+1)$-photon system is in the state
\begin{eqnarray}
|\Psi''_1\rangle_{Z\cdots CBAa_1} &=&\frac{1}{\sqrt{N}}[\vert
H\rangle_Z\cdots|H\rangle_C\vert H\rangle_B \vert V\rangle_A \vert
V\rangle_{a_1} \nonumber\\
&+&  (\vert H\rangle_Z\cdots \vert H\rangle_C \vert V\rangle_B +
\vert H\rangle_Z\cdots \vert V\rangle_C \vert H\rangle_B \nonumber\\
&+& \cdots + \vert V\rangle_Z\cdots \vert H\rangle_C \vert
H\rangle_B)\vert H\rangle_A\vert H\rangle_{a_1}],
\label{state09}
\end{eqnarray}
which takes place with the probability
\begin{eqnarray}
P''_1= \frac{N\alpha_1^2\beta_1^2}{  \alpha_1^2 + \beta_1^2 }.
\label{P}
\end{eqnarray}
If the parity of the two photons $A$ and $a_1$ is odd, the
$(N+1)$-photon system is in the state
\begin{eqnarray}
|\Psi'''_1\rangle_{Z\cdots CBAa_1} &=&\frac{1}{\sqrt{\alpha_1^4
+(N-1) \beta_1^4}} [\alpha_1^2 \vert H\rangle_Z\cdots
|H\rangle_{C}|H\rangle_{B}|V\rangle_{A}\vert
H\rangle_{a_1}   \nonumber\\
&+& \beta_1^2 (\vert H\rangle_Z\cdots|H\rangle_{C}|V\rangle_{B} +
\vert
H\rangle_Z\cdots|V\rangle_{C}|H\rangle_{B} \nonumber\\
&+& \cdots + \vert
V\rangle_Z\cdots|H\rangle_{C}|H\rangle_{B})|H\rangle_{A}|V\rangle_{a_1}
],
\label{state10}
\end{eqnarray}
which takes place with the probability
\begin{eqnarray}
P'''_1= \frac{\alpha_1^4 + (N-1)\beta_1^4}{  \alpha_1^2 + \beta_1^2
}. \label{P}
\end{eqnarray}

By measuring the ancillary photon $a_1$ in the $(N+1)$-photon system
in the state $|\Psi_1\rangle_{Z\cdots CBAa_1}$, the $N$ parties can
obtain the standard $N$-photon state
\begin{eqnarray}
|W^+_N\rangle_{Z\cdots CBA} &=&\frac{1}{\sqrt{N}}[\vert
H\rangle_Z\cdots|H\rangle_C\vert H\rangle_B \vert V\rangle_A  \nonumber\\
&+&  (\vert H\rangle_Z\cdots \vert H\rangle_C \vert V\rangle_B +
\vert H\rangle_Z\cdots \vert V\rangle_C \vert H\rangle_B \nonumber\\
&+& \cdots + \vert V\rangle_Z\cdots \vert H\rangle_C \vert
H\rangle_B)\vert H\rangle_A ]
\label{state09}
\end{eqnarray}
with or without a phase-flip operation on the photon $A$. When the
$(N+1)$-photon system is in the state $|\Psi'''_1\rangle_{Z\cdots
CBAa_1}$, Alice can collapse it into the $N$-photon state
\begin{eqnarray}
|\Psi'_1\rangle_{Z\cdots CBA} &=&\frac{1}{\sqrt{\alpha_1^4 +(N-1)
\beta_1^4}} [\alpha_1^2 \vert H\rangle_Z\cdots
|H\rangle_{C}|H\rangle_{B}|V\rangle_{A}    \nonumber\\
&+& \beta_1^2 (\vert H\rangle_Z\cdots|H\rangle_{C}|V\rangle_{B} +
\vert
H\rangle_Z\cdots|V\rangle_{C}|H\rangle_{B} \nonumber\\
&+& \cdots + \vert
V\rangle_Z\cdots|H\rangle_{C}|H\rangle_{B})|H\rangle_{A} ] 
\label{state10}
\end{eqnarray}
by measuring the ancillary photon $a_1$ with the basis $X$ and
performing a phase-flip operation or not on the photon $A$.
Moreover, the state $|\Psi'_1\rangle_{Z\cdots CBA}$ has the same
form as that of the state $|\varphi\rangle_{Z\cdots CBA}$ and can be
used as the resource in next round of concentration. By iterating
the entanglement concentration process $n$ times, the total success
probability that the $N$ parties obtain a system in a standard
$N$-photon W state from each system in a partially entangled
$N$-photon W-class state is
\begin{eqnarray}
P'(n) &=& N[ \frac{\alpha_1^2\beta_1^2}{\alpha_1^2 + \beta_1^2}+
\frac{\alpha_1^4\beta_1^4}{(\alpha_1^2 + \beta_1^2)(\alpha_1^4 + \beta_1^4)} \nonumber\\
&+& \frac{\alpha_1^{8}\beta_1^{8}}{(\alpha_1^2 +
\beta_1^2)(\alpha_1^4 + \beta_1^4)(\alpha_1^8 + \beta_1^8)} +
\cdots\nonumber\\
&+& \frac{\alpha_1^{2^n}\beta_1^{2^n}}{(\alpha_1^2 +
\beta_1^2)(\alpha_1^4 +
\beta_1^4)\cdots(\alpha_1^{2^n}+\beta_1^{2^n})}]. \label{pn}
\end{eqnarray}

\section{discuss and summary}

By far, there are no ECP for photon systems in a known W-class
state, although there are two ECPs for photon system in an unknown
W-class state \cite{W1,W2}. In fact, in a practical application of
entanglement concentration, it is not difficult for the $N$ parties
in quantum communication to obtain information about the W-class
state shared by them. They need only measure a subset of samples.
The present ECP is the first one for  a known W-class state and it
is more practical in the application in future.  Compared with other
two ECPs for W-class states \cite{W1,W2}, the present ECPs has some
 advantages. First, the present ECP requires only an $N$-photon
system in each round of entanglement concentration, not two copies
of two $N$-photon entangled systems, which decreases the difficulty
of its implementation largely. Second, only one of the $N$ parties
in quantum communication perform the local unitary operation for
reconstructing the standard W state from the W-class state and she
need only communicate the classical information to other parties for
retaining or discarding their photons, which greatly simplifies the
complication of classical communication, similar to the works for a
Bell-type state \cite{shengecp,dengecppra}. Third, it has a higher
success probability than others as its total success probability
approaches to  the limit in theory. These advantages maybe makes our
ECP more feasible than other ECPs.

In summary, we  have proposed  an ECP for nonlocal $N$-photon
systems in a partially entangled pure W-class state, resorting to
ancillary single photons and parity-check measurement based on
cross-Kerr nonlinearity. Only one of the $N$ parties in quantum
communication prepares  ancillary photons and operates the
entanglement concentration process for obtaining the standard
$N$-photon W state from each partially entangled pure W-class state.
She need only tell other parties to retain or discard their photons
in the whole entanglement concentration, which greatly simplifies
the complication of classical communication, similar to the works
for a Bell-type state \cite{shengecp,dengecppra}. Third,  it has a
higher total success probability approaching to the limit in theory
by iterating the entanglement concentration process several times.
All these advantages make our ECP more feasible than others. It
maybe have good applications in quantum communication in future.

\section*{ACKNOWLEDGMENTS}

This work is supported by the National Natural Science Foundation of
China under Grant Nos. 10974020 and 11174039,  NCET-11-0031, and the
Fundamental Research Funds for the Central Universities.

\end{document}